\documentclass[twoside,english,final,5p,twocolumn,authoryear]{elsarticle}
\usepackage[T1]{fontenc}
\usepackage[latin9]{inputenc}
\usepackage{textcomp}
\usepackage{amstext}
\usepackage{graphicx}

\makeatletter

\journal{Advances in Space Research}

\makeatother

\usepackage{babel}
\begin{document}

\begin{frontmatter}{}

\title{Reply on Comment on \textquotedbl{}High resolution coherence analysis
between planetary and climate oscillations\textquotedbl{} by S. Holm}

\author[rvt]{Nicola~Scafetta}

\ead{nicola.scafetta@unina.it}

\address[rvt]{Department of Earth Sciences, Environment and Georesources, University
of Naples Federico II, Monte Sant'Angelo, Naples, Italy.}

\address{Advances in Space Research: https://doi.org/10.1016/j.asr.2018.05.014}
\begin{abstract}
Holm (ASR, 2018) claims that Scafetta (ASR 57, 2121-2135, 2016) is
``\textit{irreproducible}'' because I would have left ``\textit{undocumented}''
the values of two parameters (a reduced-rank index $p$ and a regularization
term $\delta$) that he claimed to be requested in the Magnitude Squared
Coherence Canonical Correlation Analysis (MSC-CCA). Yet, my analysis
did not require such two parameters. In fact: 1) using the MSC-CCA
reduced-rank option neither changes the result nor was needed since
Scafetta (2016) statistically evaluated the significance of the coherence
spectral peaks; 2) the analysis algorithm neither contains nor needed
the regularization term $\delta$. Herein, I show that Holm could
not replicate Scafetta (2016) because he used different analysis algorithms.
In fact, although Holm claimed to be using MSC-CCA, for his figures
2-4 he used a MatLab code labeled ``\textit{gcs\_cca\_1D.m}'' (see
paragraph 2 of his Section 3), which Holm also modified, that implements
a different methodology known as the Generalized Coherence Spectrum
using the Canonical Correlation Analysis (GCS-CCA). This code is herein
demonstrated to be unreliable under specific statistical circumstances
such as those required to replicate Scafetta (2016). On the contrary,
the MSC-CCA method is stable and reliable. Moreover, Holm could not
replicate my result also in his figure 5 because there he used the
basic Welch MSC algorithm by erroneously equating it to MSC-CCA. Herein
I clarify step-by-step how to proceed with the correct analysis, and
I fully confirm the 95\% significance of my results. I add data and
codes to easily replicate my results.
\end{abstract}

\end{frontmatter}{}

\section{Introduction}

Although I thank Holm for his interest in my work, his critique of
\citet{Scafetta2016} is  incorrect.

\citet{Holm2017} claims that the Magnitude Squared Coherence Canonical
Correlation Analysis (MSC-CCA) by \citet{Santamaria(2007)} would
necessarily require the adoption of two additional parameters: a regularization
parameter $\epsilon$ (or $\delta$) and a reduced-rank parameter
$p$. Since \citet{Scafetta2016} did not specify their values and
he failed to reproduce my results, \citet{Holm2017} concluded that
\citet{Scafetta2016} would be ``\textit{irreproducible}.'' Thus,
he questioned my scientific results regarding the existence of a spectral
coherence (in particular at the 20- and 60-year periods) between the
global surface temperature record and the Sun's speed (SS) relative
to the solar system barycenter (see Figure 1), as first proposed in
\citet{Scafetta2010}. 

\begin{figure}
\centering{}\includegraphics[width=0.9\columnwidth]{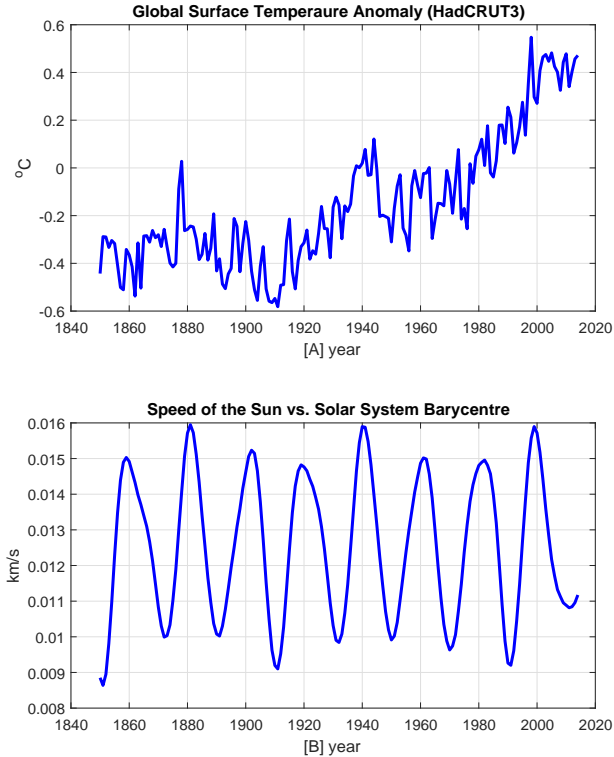}\caption{{[}A{]} HadCRUT3 global surface temperature \citep{Brohan}; {[}B{]}
Speed of the Sun relative to the barycenter of the solar system. Annual
means. \citep[For details: ][]{Scafetta2014,Scafetta2016}.}
\end{figure}

Herein I explain that \citet{Scafetta2016} did not specify any value
for such two parameters simply because my analysis did not require
them. In fact:\\
1) Holm confused the MSC-CCA method for its reduced-rank approximation
known as the Reduced-Rank CCA (MSC-RRCCA). I will explain how the
two techniques are used and how the reduced-rank parameter $p$ should
be chosen. The reduced-rank operation was developed to attempt to
suppress the \textit{noise} in order to emphasize the \textit{signal}
\citep{Santamaria(2007)}, but it was unnecessary in \citet{Scafetta2016}
since I directly evaluated the 95\% significance of the MSC-CCA spectral
peaks using the random phase significance model \citep{Traversi}.
In any case, for the specific analysis presented in \citet{Scafetta2016},
MSC-CCA and MSC-RRCCA produce identical results when $p$ is varied
within its allowed range. Therefore, \citet{Scafetta2016}'s result
could not be ambiguous.\\
2) Regarding the regularization parameter $\epsilon$ (or $\delta$),
it is evident that it was added to the algorithm by Holm himself.
In fact, this parameter simply does not exist in the original MSC-CCA/MSC-RRCCA
definition or code and, therefore, I could not have used it. Herein
I will explain why MSC-CCA/MSC-RRCCA, in most cases, does not require
it even when the correlation matrices are singular. 

Contrary to Holm's opinion, his failure to reproduce \citet{Scafetta2016}
was not due to any ambiguity present in my work regarding presumed
undocumented parameters. Holm just used different analysis algorithms
instead of the real MSC-CCA/MSC-RRCCA one. In fact, for his figures
2-4, \citet[section 3, paragraph 2]{Holm2017} apparently adopted
a MatLab ``\textit{gcs\_cca\_1D.m}'' function (modified with a regularization
parameter) that, as people familiar with these codes know, evaluates
the Generalized Coherence Spectrum using the Canonical Correlation
Analysis (GCS-CCA), proposed only in \citet{Ramirez}, which \citet{Scafetta2016}
did not even cite. For his figure 5, Holm used the basic form of the
Welch MSC algorithm (his eq. 3) by erroneously equating it to the
non-parametric MSC-CCA (his eq. 11). Thus, \citet{Holm2017} is rather
misleading since he always claimed to use the MSC-CCA methodology
while, in reality, he adopted different MSC methodologies. 

To avoid any possible misinterpretation, I now provide as an electronic
supplement the Matlab codes to replicate the MSC-CCA analysis of \citet{Scafetta2016}. 

\section{MSC-CCA and MSC-RRCCA}

MSC-CCA and MSC-RRCCA are differently defined. \citet{Santamaria(2007)}
clearly distinguished between them although the main intent of their
work was to develop the parametric MSC-RRCCA algorithm, which is one
of the improved versions of MSC-CCA. \citet{Zheng} showed that MSC-CCA
belongs to a family of non-parametric MSC estimators of the type:

\begin{equation}
^{\alpha}\gamma_{xy}^{2}(\omega_{L})=\frac{\left|\mathbf{f}_{L}^{H}\mathbf{R}{}_{xx}^{\text{\textminus}\alpha}\mathbf{R}_{xy}\mathbf{R}{}_{yy}^{\text{\textminus}\alpha}\mathbf{f}_{L}\right|^{2}}{(\mathbf{f}_{L}^{H}\mathbf{R}{}_{xx}^{1\text{\textminus}2\alpha}\mathbf{f}_{L})(\mathbf{f}_{L}^{H}\mathbf{R}{}_{yy}^{1\text{\textminus2}\alpha}\mathbf{f}_{L})},\label{eq:basic}
\end{equation}
where $x$ and $y$ are two time series of $N$ data, $\mathbf{R}_{xx}$,
$\mathbf{R}_{yy}$ and $\mathbf{R}_{xy}$ are the correlation and
cross-correlation matrices, $\mathbf{f}_{L}$ is the Fourier vector,
$\mathbf{f}_{L}=\left[1\:e^{jw}\:\ldots\:e^{jw(L-1)}\right]^{T}/\sqrt{L}$,
$L$ is the window length parameter, $\omega$ is the frequency, $0\leq{}^{\alpha}\gamma_{xy}^{2}\leq1$
and $\alpha\in[0,1]$ is the exponential characterizing the estimator.
MSC-CCA is defined as:

\begin{equation}
\gamma_{xy}^{2}(\omega,L)=\left|\mathbf{f}_{L}^{H}\mathbf{R}_{xx}^{\text{\textminus}0.5}\mathbf{R}_{xy}\mathbf{R}_{yy}^{\text{\textminus}0.5}\mathbf{f}_{L}\right|^{2},\label{eq:1}
\end{equation}
where $\mathbf{C}_{xy}=\mathbf{R}_{xx}^{\text{\textminus}0.5}\mathbf{R}_{xy}\mathbf{R}_{yy}^{\text{\textminus}0.5}$
is the coherence matrix. The adoption of the square root ($\alpha=0.5$)
makes MSC-CCA a midway algorithm between the Welch ($\alpha=0$) and
the MVDR ($\alpha=1$) MSC methods which optimizes its MSC performance.
In fact, as the parameter $\alpha$ decreases from 1 to 0, the signal
mismatch problem reduces at the expense of a decrease in frequency
resolution \citep{Zheng}. In fact, computer tests demonstrate the
MSC-CCA advantages such as a better spectral resolution versus the
Welch\textquoteright s estimator (implemented in the MatLab mscohere
function) and the avoidance of the signal canceling problems of the
minimum variance distortion-less response (MVDR) estimator \citep[cf.:][]{Santamaria(2007),Scafetta2016,Zheng}.

An optional operation can be added to the MSC-CCA algorithm to filter
out the lowest MSC value frequencies, which are interpreted as noise
or non-coherent signals. \citet{Santamaria(2007)} labeled this methodology
\textit{Reduced-Rank CCA} (MSC-RRCCA). This operation is possible
because $\mathbf{C}_{xy}$ can be decomposed by singular value decomposition
(SVD) as $\mathbf{C}_{xy}=\mathbf{U}\mathbf{\Lambda}\mathbf{U}^{H}$:
where $\mathbf{U}\mathbf{U}^{H}=\mathbf{I}$ , $\mathbf{U}$ contains
the singular vectors of $\mathbf{C}_{xy}$ and $\mathbf{\Lambda}$
is a diagonal matrix with non-negative real singular eigenvalues,
$k_{i}^{2}$, for $i=1,\ldots,\:L$, sorted in descending order. Thus,
it is possible to select a number $p<L$ of eigenvalues considered
to be the most significant ones, and substitute the coherence matrix
$\mathbf{C}_{xy}$ with its reduced-rank approximation of order $p$:
$\tilde{\mathbf{C}}_{xy,p}=\mathbf{U}\tilde{\mathbf{\Lambda}}_{p}\mathbf{U}^{H}$.
\citet[figre 4]{Santamaria(2007)} only proposed a qualitative methodology
for the choice of $p$ based on a visual inspection of how the singular
eigenvalue function drops: their examples suggests that $p$ could
be chosen as $k_{p}^{2}>0.5\geq k_{p+1}^{2}$. More recently, \citet{Shao}
proposed a generalized likelihood ratio test (GLRT) methodology. In
any case, the RR diagonal matrix $\tilde{\mathbf{\Lambda}}_{p}$ is
obtained by setting $k_{i}^{2}=0$ for $i=p+1,\ldots,\:L$, and MSC-RRCCA
is defined as 

\begin{equation}
\tilde{\gamma}_{xy}^{2}(\omega,L,p)=\left|\mathbf{f}_{L}^{H}\tilde{\mathbf{C}}_{xy,p}\mathbf{f}_{L}\right|^{2},\label{eq:1-1}
\end{equation}
with $0\leq\tilde{\gamma}_{xy}^{2}\leq\gamma_{xy}^{2}\leq1$ \citep[e.g. ][]{Shao}.
Eqs. \ref{eq:1} and \ref{eq:1-1} show that MSC-CCA and MSC-RRCCA
differ. However, when $p=L$ the latter exactly coincides with the
former and, therefore, MSC-RRCCA generalizes MSC-CCA. When $p<L$,
MSC-RRCCA is a kind of MSC-CCA filtered off of its less relevant MSC
values. However, if the only excludable singular eigenvalues are already
equal to zero, which occurs when $\mathbf{C}_{xy}$ is singular, MSC-RRCCA
and MSC-CCA produce exactly the same output. This is the case for
the analysis performed in \citet{Scafetta2016} where I used a traditional
method to directly evaluate the 95\% significance of the MSC spectral
peaks. 

Based on the above definitions it is clear that \citep{Holm2017}'s
claim that MSC-CCA \textit{``assumes a model with a predetermined
number of sinusoids for the climate data''} is erroneous. In fact,
MSC-CCA does not eliminate any singular eigenvalues of the coherence
matrix $\mathbf{C}_{xy}$ and, therefore, it does not apply any frequency
filtering or selection. Moreover, such a selection does not occur
even when MSC-RRCCA is adopted if the excluded singular eigenvalues
of $\mathbf{C}_{xy}$ are already all equal to zero.

\section{\citet{Scafetta2016} cited and used MSC-CCA}

\citet{Holm2017}'s main allegation is that \citet{Scafetta2016}
did not specify the \textit{``values of key parameters in the CCA
method''} that, in his opinion, I should have necessarily adopted.
The first one would be the reduced-rank parameter $p$ introduced
above. Yet, \citet{Scafetta2016} only used MSC-CCA in its basic form
as implicit in the fact that I did not explicit any value of $p$.
Thus, a reader had to realize that the RR option was not used or that
I used it at its default value $p=L$. Evidently, Scafetta had no
obligation to explicit the value of a parameter that is either missing
in the MSC-CCA algorithm (Eq. \ref{eq:1}) or it is automatically
set to its default value by the original code itself. 

In fact, the Matlab reduced-rank MSC-CCA code provided by its authors,
\textit{``CCA\_MSC.m,''} contains also the command ``\texttt{if
isempty(R) R=L;}'' which sets the reduced rank parameter $p$ (labeled
$R$) to $L$ when its input is left empty. To avoid any possible
confusion or misinterpretation, I now provide as an electronic supplement
the Matlab codes to replicate the MSC-CCA analysis of \citet{Scafetta2016}.
Thus, according to its own authors, MSC-CCA is the default version
of MSC-RRCCA, as the mathematical logic of the equations 1 and 2 also
imply. On the contrary, Holm's misunderstanding likely occurred because
he used the MatLab ``\textit{gcs\_cca\_1D.m}'' function, written
as \textit{gcs\_cca\_1D(x,L,K,P)} (see Supplement), that depends explicitly
on a reduced-rank parameter ``$P$'' that must be set to some value. 

Moreover, \citet{Scafetta2016} used the adjective ``\textit{reduced-rank}''
just in page 2126 when I introduced the content of \citet{Santamaria(2007)}
that compared several MSC methods, but I never used it when I presented
or discussed my own calculations. I was also very careful to title
figures 4 and 7 in \citet{Scafetta2016} just as \textit{``Canonical
Coordinates''} and \textit{``Canonical Coordinates (CCA)''}, respectively,
while \citet{Santamaria(2007)} titled their figures 1-3 \textit{``reduced-rank
CCA''} since they showed MSC-RRCCA examples while I showed MSC-CCA
ones. Note that also \citet{Zheng} showed examples of MSC-CCA without
any reduced rank.

Despite the numerous evidences that I did not use the RR option, \citet{Holm2017}
only exploited a possible minor typo present in page 2126 of \citet{Scafetta2016}
to claim that I was ambiguous regarding whether I was using MSC-RRCCA
or MSC-CCA. Yet, the typo likely occurred because the original authors
labeled their code as ``\textit{CCA\_MSC.m}.'' This label is ambiguous
since the code actually implements MSC-RRCCA while MSC-CCA is interpreted
as its default state (see Supplement). Consequently, I likely wrote
``\textit{CCA\textendash MSC is based on the reduced rank coherence
matrix...}'' because I was implicitly referring to the MatLab code
label. However, it is true that the use of the term \textquotedbl{}reduced-rank\textquotedbl{}
in \citet{Scafetta2016} might have confused a few readers. In any
case, Section 5 demonstrates that Holm's ``ambiguity argument''
is irrelevant because under the same statistical condition of the
analysis proposed in \citet{Scafetta2016}, both MSC-CCA and MSC-RRCCA
produce an identical result. 

\section{Holm's regularization parameter is unnecessary}

\citet{Holm2017}'s second claim is that Eq. \ref{eq:1} had to be
modified using a regularization parameter $\epsilon$ (or $\delta$),
which he supposed that I had used too but I left \textit{``undocumented''}.
His Eq. 9 expressed such a modification as $\mathbf{R}_{xx}=\mathbf{R}_{xx}+\epsilon\mathbf{I}$,
which is also improperly written because it would imply $\epsilon=0$,
while Holm set $\epsilon\neq0$. Evidently, \citet{Scafetta2016}
did not mention any regularization parameter simply because it does
not exist in Eq. \ref{eq:1} nor in Eq. \ref{eq:1-1}. Moreover, it
is not mentioned in \citet{Santamaria(2007)} nor included in their
``\textit{CCA\_MSC.m}'' code. 

Holm motivated the addition of such a regularization parameter because
when he tried his \textit{``gcs\_cca\_1D.m''} function on the physical
records he found MSC estimates often larger than unity. \citet{Holm2017}
interpreted his  results by claiming that a regularization parameter
would be necessary to avoid \textit{``numerical problems due to possible
singularity of''} the correlation matrices. Yet, Holm's statements
are explicit admissions only that his code was not working properly.
Indeed, MSC values cannot be larger than unity, which implies a numerical
problem or a mathematical flaw in the algorithm or in the code. 

Contrary to Holm's opinion, I found that adding a regularization parameter
to the MSC-CCA algorithm is often unnecessary because most matrix
singularity issues are already efficiently handled by MatLab when
the code is written as in the Supplement. Essentially, in processing
Eq. \ref{eq:1}, Matlab often evaluates $\mathbf{R}_{xx}^{-1/2}$
and $\mathbf{R}_{yy}^{-1/2}$ and then it factors their theoretical
infinities using the singularities of $\mathbf{R}_{xy}$ as if $\sqrt{\infty}\cdot0\cdot\sqrt{\infty}=0$,
which prevents the NaN error. The same result could be obtained by
adding a very small regularization term to the correlation matrices
but, as said, in \citet{Scafetta2016} this was unnecessary since
Matlab did not give any warnings regarding an encountered numerical
failure. In fact, simple tests show that Matlab evaluates $\mathbf{A}^{-0.5}$
even when $\mathbf{A}^{-1}$ fails because the matrix $\mathbf{A}$
is singular. Probably, the computational rounding errors slightly
break the matrix singularity and its positive semi-definite status.
Then, the square root makes it easier to keep the values within the
double floating-point limits of the computer that can handle positive
real numbers between $2\cdot10^{-308}$ and $2\cdot10^{308}$. This
fact makes the MSC-CCA code significantly more stable than, for example,
the MVDR estimator \citep{Benesty} whose published code uses a small
and fixed regularization parameter (which I did not change) to slightly
modify the correlation matrices to permit their numerical inversion
in singularity cases.

Moreover, once the coherence matrix $\mathbf{C}_{xy}$ of Eq. \ref{eq:1}
is numerically well defined, the RR operation can be applied without
any problem. Thus, also \citet{Holm2017}'s claim that the regularization
parameter would be \textit{``unnecessary'' }when the RR option is
not used\textit{,} is incorrect. It does not reflect how Eqs. \ref{eq:1}
and \ref{eq:1-1} work, which necessarily imply that $0\leq\tilde{\gamma}_{xy}^{2}\leq\gamma_{xy}^{2}\leq1$
\citep{Shao}, while Holm's claim would imply that in some cases and
for some frequencies $0\leq\gamma_{xy}^{2}\leq1<\tilde{\gamma}_{xy}^{2}$.

Regarding the GCS-CCA method, Holm's own tests (figures 2-4) showed
that it still did not properly work even after the addition of his
regularization parameter. Thus, contrary to Holm's opinion, the\textit{
gcs\_cca\_1D.m} function could not be fixed in his proposed manner.
Proposing a proper correction of the GCS-CCA method and/or of its
code is out of the scope of this work.

\section{Step-by-step replication of \citet{Scafetta2016}}

The MSC-CCA analysis (Eq. \ref{eq:1}) of the physical data is repeated
as in \citet{Scafetta2016} using $L=110$ and compared against the
MSC-RRCCA one: see Figure 2. There are $N=165$ annual values from
1850 to 2014, the last year when the HadCRUT3 temperature record by
\citet{Brohan} was provided. I used this record, and not the updated
HadCRUT4 version, because both Holm and I have used it since 2010:
the results would not change significantly using HadCRUT4. The temperature
record is detrended of its quadratic trend as done in \citet{Scafetta2016}
before the analysis.

Figure 2A shows the singular eigenvalues $k_{i}^{2}$ for $i=1,\ldots\:,110$
of the coherence matrix $\mathbf{C}_{xy}$ ordered from the larger
to the smaller. They are nearly equal to 1 for $i=1,\ldots\:,56$
and nearly equal to zero for $i=57,\ldots\:,110$. Let us make some
considerations:

1) The 56 singular eigenvalues larger than zero are expected since
the rank of $\mathbf{R}_{xx}$, $\mathbf{R}_{yy}$ and $\mathbf{R}_{xy}$
is $165-110+1=56$ while their size is $110\times110$: only 56 110-long
different moving windows can be made using 165 data.

2) The other $110-56=54$ singular eigenvalues of $\mathbf{C}_{xy}$
are nearly equal to zero. Thus, MatLab efficiently worked around the
theoretical singularities of the matrices. 

3) Figure 2A also suggests that if MSC-RRCCA is applied, the reduced-rank
parameter $p$ could be set only between 56 and 110. In fact, $p$
should not be set below 56 because all singular eigenvalues are nearly
equivalent to each other and close to 1, and it would be impossible
to discriminate them between the most and the least relevant ones.
This consideration would be also consistent with the GLRT-based rank
detection method \citep{Shao} since its test variable $2\varsigma_{p}=-N\sum_{i=p+1}^{L}\ln(1-\gamma_{i}^{2})$
would essentially be equal to infinite for $p\leq56$ and equal to
zero for $p\geq57$. Thus, within the allowed range, the selection
of $p$ does not produce any variation in the result because $k_{i}^{2}=0$
for $i=57,\ldots\:,110$. 

\begin{figure}[t]
\centering{}\includegraphics[width=0.9\columnwidth]{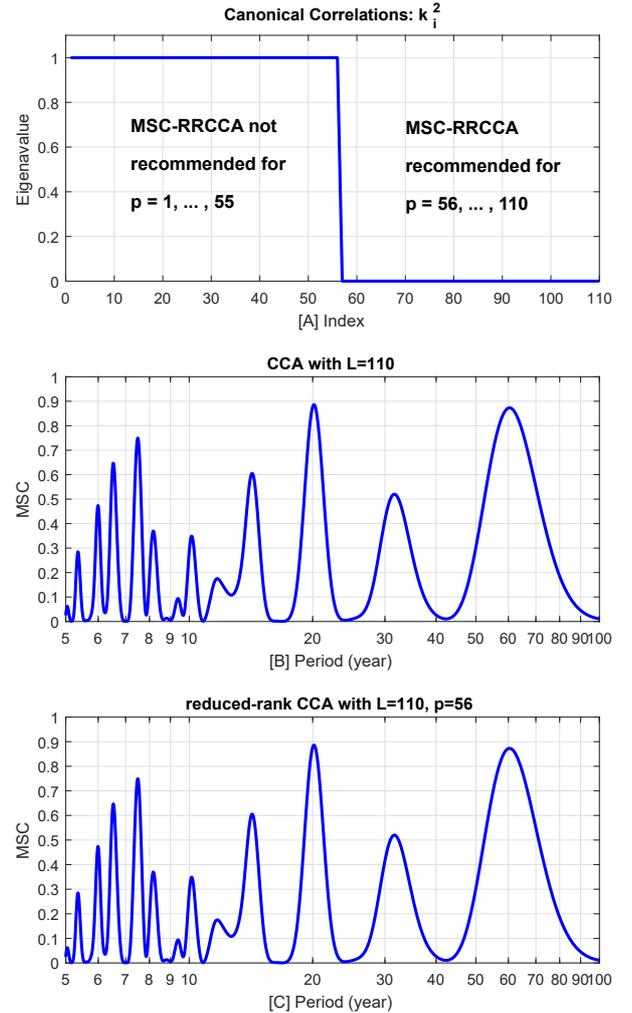}\caption{Coherence analysis between reccomended the data shown in Figure 1
using $L=110$. {[}A{]} Singular eigenvalues; {[}B{]} CCA; {[}C{]}
Reduced-Rank CCA with $p=56$. Compare with \citet{Scafetta2016}.}
\end{figure}

\begin{figure*}[t]
\centering{}\includegraphics[width=0.9\textwidth]{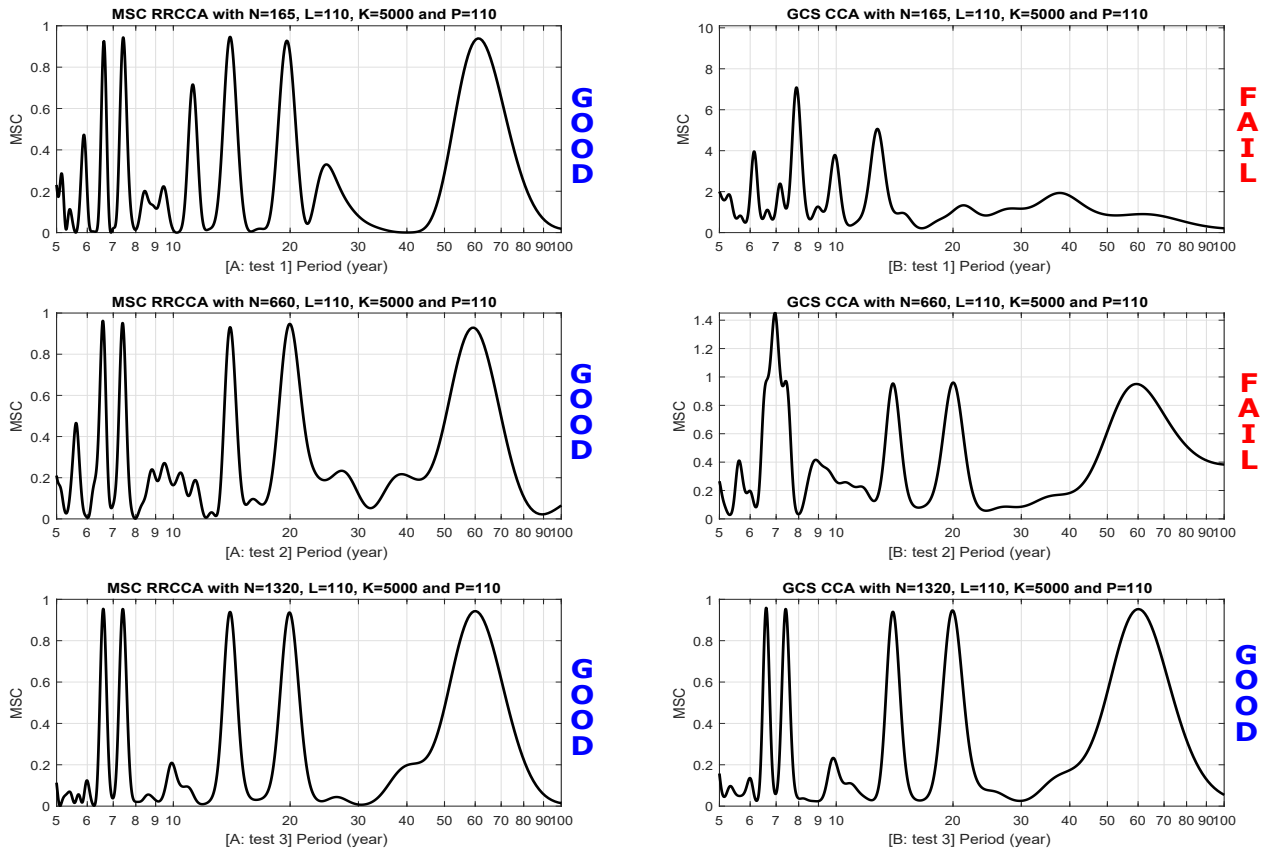}\caption{Comparison between MSC-CCA (left) and GCS-CCA (right) with the given
parameters using synthetic records as discussed in Section 6. MSC-CCA
performs always well and it is stable because the five theoretical
coherent periods equal to 6.6, 7.4, 14, 20 and 60 year are always
found. On the contrary, GCS-CCA performs well only in test \#3. Note
the severe failure of GCS-CCA in test \#1 and test \#2 since values
with $MSC>1$ are observed. }
\end{figure*}

Figures 2B and 2C show that MSC-CCA and MSC-RRCCA with $p=56$ produce
exactly the same result. Both analyses show very strong coherence
peaks at 20- and 60-year periods ($MSC\approx0.9$) and reproduce
exactly \citet[figure 4D and 7]{Scafetta2016}. 

The result is confirmed also using the monthly record as in \citet{Holm2017}
(\citet{Scafetta2016} used the annual one) and using a coherence
window $L=(N+1)/2=83$ (see also Section 8): in the latter case the
matrices $\mathbf{R}_{xx}$, $\mathbf{R}_{yy}$ and $\mathbf{R}_{xy}$
are not singular and, therefore, there are no numerical issues. The
Supplement also includes the original ``\textit{CCA\_MSC.m}'' code
to reproduce \citet{Santamaria(2007)} to assure readers that I am
using the right algorithm. Moreover, the reliability of my MatLab
codes is further confirmed by a simple computer experiment simulating
the same statistical constrains of the physical data analyzed in \citet{Scafetta2016}:
see also Figure 3. 

In conclusion, contrary to \citet{Holm2017}'s claims, \citet{Scafetta2016}
cannot be ambiguous because (1) the regularization parameter $\delta$
was not required and (2) both MSC-CCA and MSC-RRCCA produce the same
output. A reader simply had to use the indicated analysis techniques
(MSC-CCA or MSC-RRCCA would have been equivalent) and to do it properly,
but Holm used different algorithms. Then, \citet[figure 7]{Scafetta2016}
directly evaluated the significance of the MSC-CCA results using Monte
Carlo simulations based on the random phase model: see also Section
8.

\section{MSC-CCA versus GCS-CCA}

\citet{Ramirez} apparently considered the GCS-CCA as a multi-sequence
extension of MSC-CCA, which processes just two sequences \citep{Santamaria(2007)}.
However, I will herein demonstrate that this is not the case. The
two codes can generate significantly different results under specific
statistical conditions, which explains the alternative conclusions
in \citet{Scafetta2016} and \citet{Holm2017}. In fact, the \textit{``gcs\_cca\_1D.m''}
function was not written in such a way to naturally implement the
MSC-CCA algorithm in a 2-signal case because the two functions handle
the data differently. For example, for two records GCS-CCA uses 2Lx2L
coherence matrices while MSC-CCA uses LxL coherence matrices. 

Since \citet{Holm2017} found that the \textit{gcs\_cca\_1D.m} function
was unreliable using specific physical data while in \citet{Ramirez}
it was working well using generic synthetic examples, it is necessary
to test whether and under which specific statistical circumstances
GCS-CCA and MSC-CCA give different results. I will do this now by
comparing simple computer tests where the same pair of synthetic records
are processed with both techniques. I used the \textit{gcs\_cca\_1D.m}
function that the authors sent me in 2014, which replicates \citet{Ramirez}
(see Supplement). However, this function might not coincide with that
used in \citet{Holm2017} because (1) Holm did not published it and
(2) he stated that, since it was not working, he and/or Ram\'irez
altered it with a regularization parameter $\delta$. The exact nature
of the code modifications were not provided in \citet{Holm2017} so
that also his figures 2-4 cannot be replicated.  Therefore, my results
might appear different from those that Holm could get with his code,
but they show what the original \textit{gcs\_cca\_1D.m} function by
\citet{Ramirez} really does.

Pairs of synthetic records were generated with five harmonics with
periods equal to 6.6, 7.4, 14, 20 and 60 year, as similarly found
in the physical data discussed in \citet{Scafetta2016}, plus Gaussian
noise: see the Supplement for details. I use $L=110$, $K=5000$ equispaced
frequencies and $P=L=110$, which means that the reduced-rank option
was not used. However, I progressively increase the length of the
record as $N_{1}=165$ (as in the original physical records), $N_{2}=4*165$
and $N_{3}=8*165$.

Figure 3 depicts the results. MSC-CCA works always well in all three
tests, it is stable and always finds the expected 5 coherent harmonics,
which are characterized by $0.9<MSC<1$. On the contrary, GCS-CCA
works well only when N is very large relative to $L$ (test \#3) but,
as $N$ decreases, it becomes progressively more and more unstable
and completely fails for $N=165$ where the\textit{ ``estimate of
MSC often became much higher than unity'',} as \citet{Holm2017}
stated to have found in his tests. By running again and again the
same code, only the GCS-CCA result depicted in test \#3 remains stable,
while the result depicted in test \#1 changes greatly at each run
and always fails while that of test \#2 fails in some case. 

Thus, MSC-CCA and GCS-CCA perform similarly only when $N\gg L$. In
these simulations, GCS-CCA works well when $N\geq5L$. However, for
the specific analysis presented in \citet{Scafetta2016}, which required
investigating the low-frequency spectrum and used $L=110$ and $N=165$,
GCS-CCA fails. In conclusion, since GCS-CCA statistically collapses
when it attempts to reproduce \citet{Scafetta2016}, using GCS-CCA
instead of MSC-CCA definitely explains Holm's inability to reproduce
my result.

\section{\citet{Holm2017}'s figure 5 is misleading}

Regarding his figure 5, \citet{Holm2017} claimed to be using MSC-CCA
with $\delta=0$ and $P=L$ (as \citet{Scafetta2016} could have done),
but I found that his result was not reproducible even when the original
\textit{gcs\_cca\_1D.m} function was used. His figure 5 shows MSC
values between 0 and 1, but I found much-higher-than-unity MSC values:
see Figure 4. I used monthly records as in \citet{Holm2017}, but
a similar failure occurs using the yearly record. Indeed, the result
depicted in Holm's figure 5 shows contradictions by his own acknowledgment
(see his Section 2) that using the original algorithm \textit{``the
estimate of MSC often became much higher than unity.''} This was
the reason why Holm added the regularization parameter $\delta$ to
the algorithm. He also observed that for high value of $p$ (note
that $p=L$ is its maximum value) $\delta$ was necessary and had
to be high since, if it was set too low (note that $\delta=0$ is
its lowest value), his algorithm gave $MSC>1$ (cf. his figure 2 and
related comments). Thus, the severe numerical instability disclosed
by Holm is perfectly consistent with my analysis, but not with the
result shown in his figure 5. 

Indeed, Holm's figure 5 was apparently not obtained by setting $\delta=0$
and $P=L$ in the same code used for his figures 2-4. Holm stated
that he used his eq. 3 that, however, is the Welch\textquoteright s
averaged periodogram method implemented in the MatLab \textquotedblleft mscohere\textquotedblright{}
function and can be obtained also from Eq. \ref{eq:basic} with $\alpha=0$.
\citet{Holm2017} justified such a choice by claiming that his eq.
3 represented MSC-CCA without rank reduction (his eq. 11), which is
erroneous \citep[cf. Section 2 and][]{Zheng}. 

\begin{figure}
\centering{}\includegraphics[width=0.9\columnwidth]{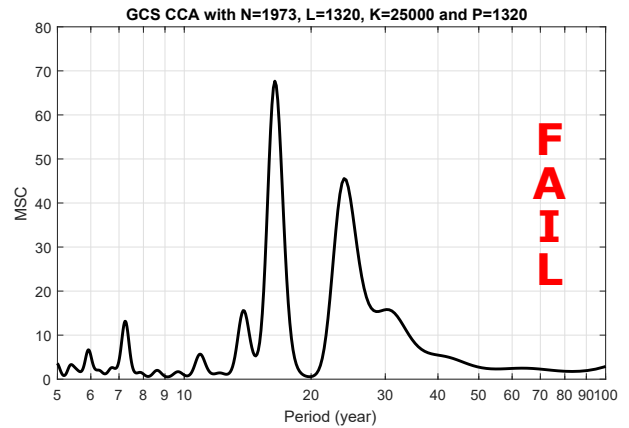}\caption{GCS-CCA analysis (with the original \textit{``gcs\_cca\_1D.m''}
function) of the physical records (monthly resolution) without the
RR filtering ($P=L$). Note the severe algorithm failure ($MSC>1$).}
\end{figure}

\begin{figure*}[t]
\centering{}\includegraphics[width=0.88\textwidth]{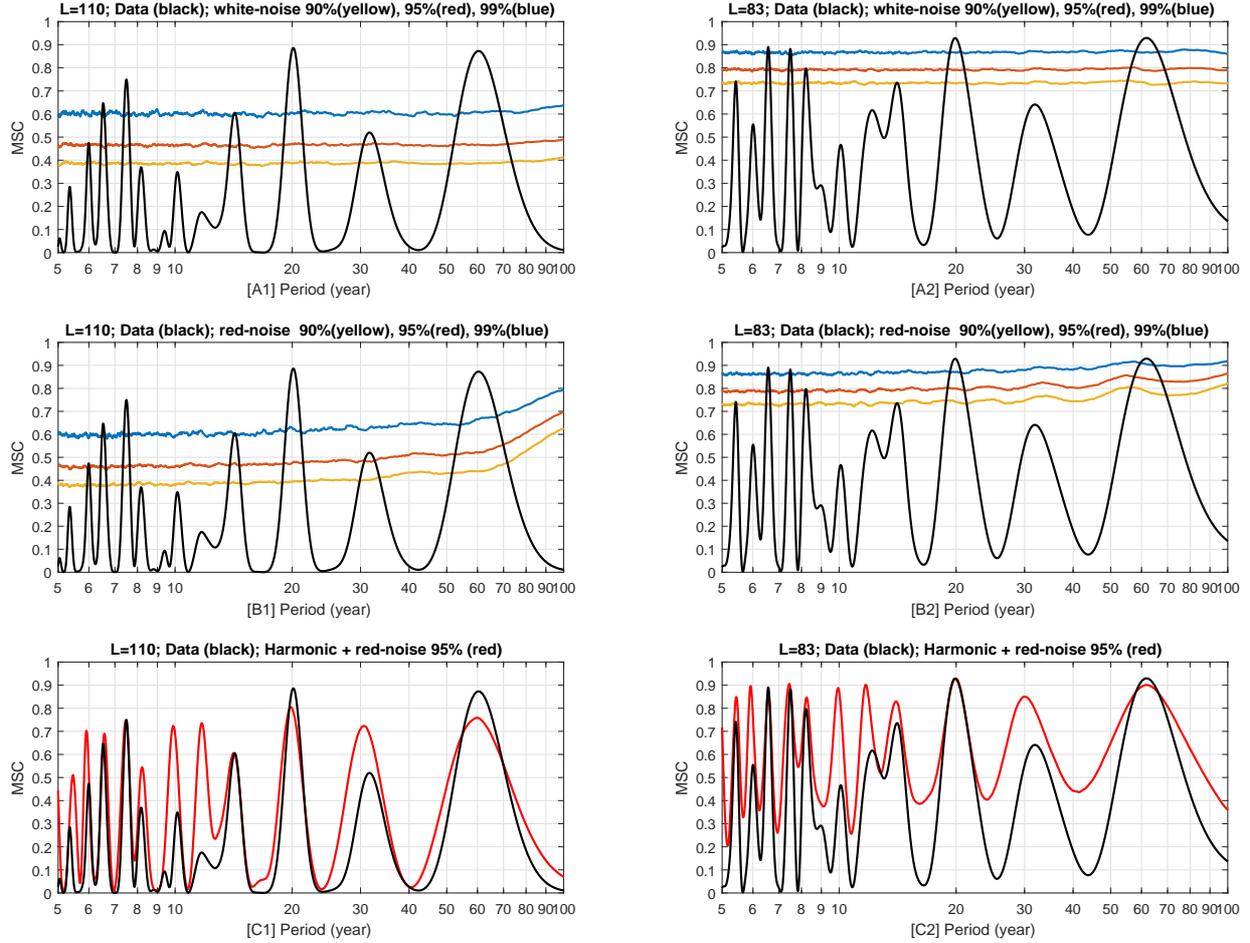}\caption{MSC-CCA between the global surface temperature record and the Sun's
speed against different significance models as explained in Section
8. The coherence frequency peaks at 20- and 60-year periods pass well
the 95\% level in all cases. }
\end{figure*}

\section{The 95\% significance problem}

\citet[figure 5]{Holm2017} also questioned that the spectral coherence
at the 20- and 60-year periods was 95\% significant. However, in \citet[figure 7]{Scafetta2016}
their 95\% significance is well met. Holm and I presumably used the
same random phase significance model \citep[cf.: ][]{Traversi} but,
as proven above, we did not use the same MSC algorithm and \citet{Scafetta2016}
already proved that the Welch MSC algorithm, which was used by Holm
instead of MSC-CCA, provides uncertain results. Moreover, there is
some issue regarding the appropriate significance model to be used.
For example, the wavelet transform coherence (WTC) proposed by \citet{Grinsted},
where AR(1) significance models are assumed by default, shows that
in the critical 17-22 year and around the 60-year range the spectral
coherence is large enough to pass well the 95\% significance level.
However, \citet[figure 5]{Holm2015} claimed a different result using
the random phase significance model, which assumes that one record
is nearly harmonic: in this case the problem could have be induced
by the WTC low spectral resolution yielding \citet{Scafetta2016}
to propose the adoption of the MSC-CCA high resolution method. Thus,
the important role of the significance model needs to be now clarified. 

The basic assumption is that two records are of the type: $x(t)=signal(t)+noise(t)$.
The issue is to determine how likely the observed MSC spectral peaks
could be artifacts of the noise function. Using Monte Carlo method
simulations, it is possible to determine the MSC-CCA significance
curves that various forms of noise models could produce. Note that
the direct adoption of the confidence methodology voids the necessity
of using the RR operation. There are three basic cases:\\
1) White-noise vs. white-noise. It is assumed that the two records
are affected just by random white noise. The test is performed by
generating 10000 pairs of Gaussian noise sequences with N=165 and
their MSC-CCA curves are evaluated. Then, for each frequency, I took
the 90\%, 95\% and 99\% top values among the 10000 estimates.\\
2) Red-noise vs. red-noise. It is assumed that both records are AR(1)
processes. The test is performed by generating 10000 pairs of AR(1)
sequences calibrated on the data records with N=165 and evaluate their
MSC-CCA curves. The AR(1) records are obtained with the model $x_{n}=\alpha x_{n-1}+\xi_{n}$
for $n=1,2,\ldots,\,N$, where $\xi_{n}$ is a sequence of Gaussian
random noise and the AR(1) parameter $\alpha$ is measured on the
physical data record. I got $\alpha=0.94$ for the astronomical record
once detrended of its mean and $\alpha=0.58$ for the temperature
record once detrended of a parabolic trend. Then, I did as above.\\
3) Harmonic signal vs. red-noise. One record is assumed harmonic while
the other (e.g. the temperature record) is an AR(1) process. This
case is interesting because a nearly harmonic record (e.g. the astronomical
one) would a-priory select specific harmonics that could give origin
to high specific MSC peaks even if tested against just noise. The
test is performed by pairing the astronomical record with 10000 AR(1)
sequences modeling the temperature record as above. Then, for each
frequency, I took the 95\% top value among the 10000 estimates.

Figure 5 shows the evaluated significance curves against the measured
MSC-CCA curves generated by the physical records using $L=110$ and
$L=83$. The noise models \#1 and \#2 give a 99\% significance for
many MSC peaks including those at 20- and 60-year periods. Test \#3
shows results similar to those found in \citet[figure 7]{Scafetta2016}
using the random phase significance model, which the method approximately
simulates, and give a very safe 95\% significance level for the same
coherence spectral peaks. Thus, the 6 panels of Figure 4 fully confirm
\citet{Scafetta2016} and my previous studies, and contradict the
contrary claims made in \citet{Holm2014,Holm2015,Holm2017}. 

Regarding test \#3, I note that if one of the two records is already
known to be harmonic \citep[e.g. orbital astronomical records, cf.:][figure 5]{Scafetta2014}
and its main harmonics are already known, using spectral coherence
methodologies should be unnecessary in most cases. In fact, in such
situations the spectral coherence is logically reduced to the simple
verification of whether the second record (e.g. the temperature one)
is characterized by spectral frequencies consistent with those already
known to exist in the harmonic signal. Spectral analysis confirms
with a 99\% significance the presence of quasi 20- and 60-year harmonics
in the global surface temperature: this was the original logic followed
in \citet[figures 3, 6 and 9]{Scafetta2010} and in \citet[figure 2B]{Scafetta2016}.

\section{Conclusion}

I have demonstrated that \citet{Holm2017} failed to reproduce \citet{Scafetta2016}
not because I left two parameters, $\delta$ and $p$, \textit{``undocumented''},
as he claimed, but because he mistook the MSC-CCA method \citep{Santamaria(2007)},
which is what \citet{Scafetta2016} used and referenced, for two different
MSC methodologies. For his figures 2-4, Holm apparently adopted the
GCS-CCA methodology proposed in \citet{Ramirez} altered with a regularization
term without realizing that it implemented a different algorithm.
Herein I showed (1) that the MSC-CCA methodology did not need the
two parameters proposed by Holm and (2) that the \textit{``gcs\_cca\_1D.m''}
function, which Holm adopted, becomes progressively unstable and collapses
under the specific statistical conditions required to replicate \citet{Scafetta2016}.
Of course, \citet{Scafetta2016} was not responsible about \citet{Ramirez},
the \textit{``gcs\_cca\_1D.m''} function and Holm adopting it to
try to reproduce my results because I always cited and used only the
reliable MSC-CCA code sent me by V\'ia. For his figure 5, Holm was
supposed to use MSC-CCA with $\delta=0$ and $P=L$ and mentioned
that he still could not replicate my results. Yet, for this figure
he used his eq. 3, representing the basic Welch MSC algorithm (Eq.
\ref{eq:basic} with $\alpha=0$) by erroneously equating it to the
non-parametric MSC-CCA algorithm (Eq. \ref{eq:basic} with $\alpha=0.5$).

Moreover, no formal ambiguity could exist in \citet{Scafetta2016}
regarding the adoption of the RR parameter $p$, as Holm also charged,
because in the specific case both MSC-CCA and MSC-RRCCA produce the
exact same result when properly used. Moreover, the RR option was
not needed in \citet{Scafetta2016} because I adopted Monte Carlo
simulations based on the random phase model to evaluate the 95\% statistical
significance of MSC spectral peaks. Thus, the ``noise'' present
in the MSR-CCA result did not need to be suppressed with a RR filtering.
Finally, contrary to \citet{Holm2017}'s claims, I have further confirmed
the spectral coherence with at least a 95\% significance at the 20-
and 60-year periods between the analyzed climatic and astronomical
records using various standard noise models. 

\citet{Holm2017} made secondary comments referring also to his past
critiques \citep{Holm2014,Holm2015} to my previous studies, for example
in his Section 4.3. Interested readers can find my past rebuttals
in \citet{Scafetta2014,Scafetta2016}. A latest general review on
the topic of an astronomical origin of climate oscillations throughout
the Holocene is found in \citet{Scafetta2016-1} and in its references. 

The online Supplement provides data and Matlab codes necessary to
reproduce all results shown above, the original \textit{``CCA\_MSC.m''}
and\textit{ ``gcs\_cca\_1D.m''} codes and a code to reproduce Holm's
figure 5 using the \textit{mscohere} function. See: https://doi.org/10.1016/j.asr.2018.05.014


\begin{thebibliography}{00}
\bibitem[Benesty et al.(2006)]{Benesty}Benesty, J., Chen, J., Huang,
Y. 2006. Estimation of the coherence function with the MVDR approach.
2006 IEEE Int. Conf. Acoust. and Speech Sign. Proc., 3, 500\textendash 503.

\bibitem[Brohan et al.(2006)]{Brohan}Brohan, P., Kennedy, J.J., Harris,
I., Tett, S.F.B., Jones, P.D., 2006. Uncertainty estimates in regional
and global observed temperature changes: a new dataset from 1850.
J. Geophys. Res., 111, D12106.

\bibitem[Grinsted et al.(2004)]{Grinsted}Grinsted, A., Moore, J.
C., Jevrejeva, S., 2004. Application of the cross wavelet transform
and wavelet coherence to geophysical time series. Nonlinear Proc.
Geophys., 11, 561\textendash 566.

\bibitem[Holm(2014)]{Holm2014}Holm, S., 2014. On the alleged coherence
between the global temperature and the sun\textquoteright smovement.
J. Atmos. Solar-Terrestr. Phys., 110\textendash 111, 23\textendash 27.

\bibitem[Holm(2015)]{Holm2015}Holm, S., 2015. Prudence in estimating
coherence between planetary, solar and climate oscillations. Astrophys.
Space Sci., 357, 1\textendash 8.

\bibitem[Holm(2018)]{Holm2017} Holm, S., 2018. Comment on ``High
resolution coherence analysis between planetary and climate oscillations\textquotedblright .
Advances in Space Research. https://doi.org/10.1016/j.asr.2017.09.034 

\bibitem[Ram\'irez et al.(2008)]{Ramirez}Ram\'irez, D., V\'ia,
J., Santamar\'ia, I., 2008. A generalization of the magnitude squared
coherence spectrum for more than two signals: definition, properties
and estimation. 2008 IEEE Int. Conf. Acoust. and Speech Sign. Proc.,
3769\textendash 3772.

\bibitem[Santamar\'ia and V\'ia(2007)]{Santamaria(2007)}Santamar\'ia,
I., V\'ia, J., 2007. Estimation of the magnitude squared coherence
spectrum based on reduced- rank canonical coordinates. 2007 IEEE Int.
Conf. Acoust. and Speech Sign. Proc., 3, 985\textendash 988. 

\bibitem[Scafetta(2010)]{Scafetta2010}Scafetta, N., 2010. Empirical
evidence for a celestial origin of the climate oscillations and its
implications. J. Atmos. Sol. Terr. Phys., 72, 951\textendash 970.

\bibitem[Scafetta(2014)]{Scafetta2014}Scafetta, N., 2014. Discussion
on the spectral coherence between planetary, solar and climate oscillations:
a reply to some critiques. Astrophys. Space Sci. 354, 275\textendash 299.

\bibitem[Scafetta(2016)]{Scafetta2016}Scafetta, N., 2016. High resolution
coherence analysis between planetary and climate oscillations. Advances
in Space Research, 57, 2121\textendash 2135.

\bibitem[Scafetta et al.(2016)]{Scafetta2016-1}Scafetta, N., Milani,
F., Bianchini, A., Ortolani, S., 2016. On the astronomical origin
of the Hallstatt oscillation found in radiocarbon and climate records
throughout the Holocene. Earth-Science Reviews, 162, 24\textendash 43.

\bibitem[Shao et al.(2014)]{Shao}Shao, Q., Peng, R., Zheng, C., 2014.
Estimation of a generalized non-parametric magnitude squared coherence
spectrum using the GLRT-based rank detection. In IEEE Inter. Conf.
on Signal Process. (ICSP), 189-193.

\bibitem[Traversi et al.(2012)]{Traversi}Traversi, R., Usoskin, I.,
Solanki, S., Becagli, S., Frezzotti, M., Severi, M., Stenni, B., Udisti,
R., 2012. Nitrate in polar ice: a new tracer of solar variability.
Sol. Phys., 280, 237\textendash 254.

\bibitem[Zheng et al.(2008)]{Zheng}Zheng, C., Zhou, M., Li, X., 2008.
On the relationship of non-parametric methods for coherence function
estimation. Signal Process., 88, 2863\textendash 2867.
\end{thebibliography}
\end{document}